\documentclass[conference]{IEEEtran}
\usepackage{etoolbox}
\usepackage{cite}
\usepackage{amsmath,amssymb,amsfonts}
\usepackage{algorithmic}
\usepackage{graphicx}
\usepackage{textcomp}
\usepackage{xcolor,soul}
\usepackage{enumitem}
\newtheorem{definition}{Definition}
\usepackage{dsfont}
\usepackage[ruled,vlined]{algorithm2e}
\usepackage{listings}
\usepackage{hyperref}
\usepackage{tikz}
\usepackage{pgfplots}
\pgfplotsset{compat=1.17} 
\usepgfplotslibrary{groupplots}
\usetikzlibrary{backgrounds,calc,shadings,shapes,arrows,arrows.meta,shapes.arrows,shapes.symbols,patterns, datavisualization, datavisualization.formats.functions, matrix}

\newtheorem{proposition}{\textbf{Proposition}}

\AfterEndPreamble{
  \mathchardef\mathcomma\mathcode`\,
  \mathcode`\,="8000 
}
{\catcode`,=\active
  \gdef,{\mathcomma\discretionary{}{}{}}
}
\IEEEoverridecommandlockouts
\def\BibTeX{{\rm B\kern-.05em{\sc i\kern-.025em b}\kern-.08em
    T\kern-.1667em\lower.7ex\hbox{E}\kern-.125emX}}
\begin{document}

\title{Software defined networking flow admission and routing under minimal security constraints}

\author{\IEEEauthorblockN{Jorge L\'opez, Charalampos Chatzinakis, Marc Cartigny and Claude Poletti}
\IEEEauthorblockA{\textit{Airbus}\\
Issy-Les-Moulineaux, France \\
\{jorge.lopez-c,charalampos.chatzinakis,marc.cartigny,claude.poletti\}@airbus.com}
}

\maketitle

\begin{abstract}
In recent years, computer networks and telecommunications in general have been shifting paradigms to adopt software-centric approaches. Software Defined Networking~(SDN) is one of such paradigms that centralizes control and intelligent applications can be defined on top of this architecture. The latter enables the definition of the network behavior by means of software. In this work, we propose an approach for Flow Admission and Routing under Minimal Security Constraints~(FARSec) in Software Defined Networks, where network flows must use links which are at least as secure as their required security level. We prove that FARSec can find feasible paths that respect the minimum level of security for each flow. If the latter is not possible FARSec rejects the flow in order not to compromise its security. We show that the computational complexity of the proposed approach is polynomial. Experimental results with semi-random generated graphs confirm the efficiency and correctness of the proposed approach. Finally, we implement the proposed solution using OpenFlow and ONOS -- an SDN open-source controller. We validate its functionality using an emulated network with various security levels. 
\end{abstract}


\section{Introduction}\label{sec:intro}
%global motivation software and nets, also new schemes per link
Traditionally, computer networks used dedicated hardware appliances, that fulfilled specific network functions (e.g., a firewall to filter hazardous network packets, or a network router to properly forward network packets to the appropriate next hop, etc.). Currently, computer networks are shifting this device-based paradigm. Network functions are provided as software packages or virtualized appliances. Moreover, networks are becoming generally virtualized and being executed in interconnected computer clusters. This provides ease of creation, modification, and general maintenance tasks. When it comes to the management of network traffic, paradigms have also been shifting in recent years. Notably, the Software Defined Networking~(SDN) paradigm allows the control of traffic to be decoupled from the forwarding devices. This allows one to define the behavior of the network by any means of software technologies. The latter opens up a plethora of possibilities when it comes to different ways to steer network traffic.

%particular motivation -- what about all the existing but, not addressed security, also 
With the rise of quantum communications, real implementations, and particularly, the Quantum Key Distribution~(QKD), new ways of securing communications are possible. Particularly, QKD allows secure communication \cite{seccomqkd} without relying in computational hardness as means of guaranteeing security; rather, QKD harnesses the physical properties of quantum mechanics guaranteeing that the channel cannot be read without producing noticeable perturbations. Additionally, the rise of quantum computation has also motivated the creation of different security schemes, notably the Post-Quantum Cryptography~(PQC). PQC is considered to be an end-to-end encryption scheme \cite{postquantumcrypto,psc_johana}, however, PQC can be used in a per-link encapsulation basis, specially for transit networks, operators, or meta-operators. On the one hand, considering the novel encryption schemes and new software-centric networks it is easy to consider a network routing that steers flows through secure paths, depending the appropriate application requirements. On the other hand,  best-effort routing has long been the standard routing paradigm for the Internet. However, the continuous growth of the traffic has called for a new paradigm that takes Quality of Service~(QoS) or Quality of Experience~(QoE) into consideration.

%the problem we solve
In this paper, we propose an algorithm that searches for feasible paths using the minimum security level of a link as the constraint. The minimum security level is a policy-imposed rather than an QoS-imposed requirement; the routing decisions have to comply with some Service Level Agreement~(SLA). Our framework works under the software defined networking paradigm (for preliminary concepts on SDN please refer to Section~\ref{sec:sdn}), where a centralized controller has a complete view of the network topology and is responsible for calculating the paths and installing the forwarding rules in the network devices. Furthermore, we assume that the network has different levels of security on its links. Links can be insecure (unencrypted) or use an encryption scheme (e.g., IPSec, QKD, PQC, etc.) that will define their security level. Every incoming flow in the network has a minimum security requirement. The solution searches for a feasible path where every link in the path has a security level higher or equal to the flow's requirement. If a path cannot be found the flow is rejected. The proposed solution runs in polynomial time w.r.t. the number of nodes and the number of flows in the networks. Simulated instances confirm the efficiency of the proposed method, and furthermore, its pertinence by assigning only paths that respect the security constraints (see Section~\ref{sec:experiments}). Furthermore, we implement the solution by coupling the proposed algorithm with a custom SDN service developed as an ONOS\cite{onos} (one of the leading open-source SDN controllers) application. The service monitors a network topology comprised of OpenFlow\cite{openflow} switches and upon reception of events (e.g. network flows, link security modified etc.) it queries the FARSec engine to calculate the appropriate secure paths (based on their minimal security requirements and the topology security levels). It subsequently creates the the calculated data paths by installing the appropriate OpenFlow rules on the respective devices. We showcase the overall functionality with a video demonstrator (see Section~\ref{sec:demo}).    %I joined the results and the particular motivation as there was few of both

%we can leave the narrated table of contents cuz we have space so far

The remainder of this paper is structured as follows. Section~\ref{sec:prelim} presents the necessary background, notations and definitions. Section~\ref{sec:farsec} formally describes the problem of interest, proposes a non-naive solution, and assesses its computational complexity, while Section~\ref{sec:experiments} presents the experimental evaluation of the proposed solution and illustrates the obtained results. Section ~\ref{sec:imp} describes an open-source implementation of the proposed approach. Section~\ref{sec:relwork} presents the related work, and finally, Section~\ref{sec:conclusion} concludes the paper, and presents future avenues.

\section{Preliminaries}\label{sec:prelim}
\input{figures/TIKZ_network_drawing}
\subsection{Software Defined Networking}\label{sec:sdn}
\begin{figure}[!htb]
    \centering
    \begin{tikzpicture}
    %draw data-plane
    \node[server]                               (h1)    {};
    \node               at ([xshift=-1cm]h1)    (h1l)   {$h_1$};
    \node[l3 switch]    at ([xshift=2cm]h1)     (s1)    {};
    \node               at ([yshift=1cm]s1)     (s1l)   {$r_1$};
    \node[l3 switch]    at ([xshift=2cm]s1)     (s2)    {};
    \node               at ([yshift=1cm]s2)     (s2l)   {$r_2$};
    \node[server]       at ([xshift=2cm]s2)     (h2)    {};
    \node               at ([xshift=.7cm]h2)    (h2l)   {$h_2$};
    \node[l3 switch]    at ([yshift=-1.5cm]s1)  (s3)    {};
    \node               at ([yshift=-1cm]s3)    (s3l)   {$r_3$};
    \node[server]       at ([xshift=-2cm]s3)    (h3)    {};
    \node               at ([xshift=-1cm]h3)    (h3l)   {$h_3$};
    \node[l3 switch]    at ([xshift=2cm]s3)     (s4)    {};
    \node               at ([yshift=-1cm]s4)    (s4l)   {$r_4$};
    \node[server]       at ([xshift=2cm]s4)     (h4)    {};
    \node               at ([xshift=.7cm]h4)    (h4l)   {$h_4$};
        
    %connect hosts to switches
    \draw[thick] (h1)--(s1);
    \draw[thick] (h2)--(s2);
    \draw[thick] (h4)--(s4);
    \draw[thick] (h3)--(s3);
        
    %connect the switches with an "almost" mesh topology
    \draw[thick] (s1.east)--(s2.west);
    \draw[thick] (s3.east)--(s4.west);
    \draw[thick] (s1.south)--(s3.north);
    \draw[thick] (s2.south)--(s4.north);
    \draw[thick] (s2.south west)--(s3.north east);
        
    %draw a dashed box over the data-plane
    \node[rectangle,draw,dashed,fill=none,text height=3.5cm,text width=8cm, minimum width=8cm,minimum height=4cm] at ([xshift=.85cm, yshift=-.8cm]s1) (main)  {\textbf{Data-plane}};
        
    %draw some data paths
    %path 1, h1 s1 s2 h2
    \draw[-Latex,very thick, gray!80, shorten <=0.2cm,shorten >=0.2cm]([yshift=0.15cm,xshift=0.3cm]h1.east)-- ([yshift=0.15cm]s1.west);
    \draw[-Latex,very thick, gray!80, shorten <=0.1cm,shorten >=0.1cm]([yshift=0.15cm,xshift=0.1cm]s1.east)-- ([yshift=0.15cm]s2.west);
    \draw[-Latex,very thick, gray!80, shorten <=0.3cm,shorten >=0.7cm]([yshift=0.15cm]s2.east)-- ([yshift=0.15cm]h2.west);
        
    %path 2, h1 s1 s3 s2 h2
    \draw[-Latex, dashed, very thick, gray!80, shorten <=0.2cm,shorten >=0.2cm]([yshift=-0.15cm,xshift=0.3cm]h1.east)-- ([yshift=-0.15cm]s1.west);
    \draw[-Latex,dashed, very thick, gray!80, shorten <=0.1cm,shorten >=0.3cm]([yshift=-0.15cm, xshift=-0.15cm]s1.west)-- ([yshift=-0.15cm, xshift=-0.15cm]s3.west);
    \draw[-Latex,dashed, very thick, gray!80, shorten <=0.3cm,shorten >=0.1cm]([yshift=-0.15cm]s3.north east)-- ([yshift=-0.15cm]s2.south west);
    \draw[-Latex, dashed, very thick, gray!80, shorten <=0.3cm,shorten >=0.7cm]([yshift=0.15cm]s2.east)-- ([yshift=0.15cm]h2.west);

    %%draw the SDN controller
    \node[server]       at ([yshift=2cm,xshift=1cm]s1)     (ctrl)    {};
    \node               at ([xshift=-2cm]ctrl)  (ctrll)   {SDN controller};
        
    %connect all switches with the controller
    \draw[thick, dotted, gray] (s1.north)--(ctrl.south east);
    \draw[thick, dotted, gray] (s2.north)--(ctrl.south west);
    \draw[thick, dotted, gray] ([xshift=0.1cm]s3.north east)--(ctrl.south);
    \draw[thick, dotted, gray] (s4.north west)--(ctrl.south);
        
    %draw applications
    \node[rectangle,pattern=horizontal lines, minimum width=0.5cm, minimum height=1cm]    at([yshift=1.5cm, xshift=-2cm]ctrl)    (app1)  {};
    \node[] at ([xshift=-1cm]app1)    (app1l) {App. 1};
    \node[] at ([xshift=2cm]app1)   (dots)  {\huge\ldots};
    \node[rectangle,pattern=horizontal lines, minimum width=0.5cm, minimum height=1cm]    at([xshift=2cm]dots)    (app2)  {};
    \node[] at ([xshift=1cm]app2)    (app2l) {App. $N$};
        
    connect applications to ctrl 
    \draw[thick, dotted, gray] (app1.south)--(ctrl.north);
    \draw[thick, dotted, gray] (app2.south)--(ctrl.north);
\end{tikzpicture}
    \caption{Example SDN architecture}\label{fig:sdn_ex}
\end{figure}
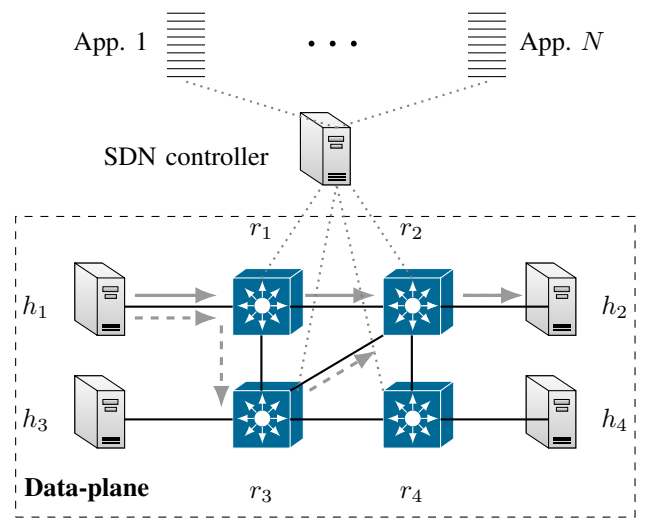

In traditional networks, the configuration, management, and data-forwarding interfaces are distributed / located at each of the data forwarding devices in the data-plane. The data-paths in the network are the result of the configuration on each of the forwarding devices; each of the devices has a local configuration and management interface. Thus, in order to re-configure the data-paths, several devices must be re-configured; as a consequence, while re-configuring each device the network may be in an inconsistent state, the process can be error-prone and slow. As an example, assume a data-plane in a traditional network as shown in Fig.~\ref{fig:sdn_ex}. Assume all flows from $h_1$ to $h_2$ follow the data-path depicted in solid arrows ($h_1\rightarrow r_1\rightarrow r_2\rightarrow h_2$); consider the link $(r_1,r_2)$ is highly loaded. In order to re-configure some of the traffic to use an alternative data-path, for example $h_1\rightarrow r_1\rightarrow r_3\rightarrow r_2\rightarrow h_2$ (depicted in dashed arrows in Fig.~\ref{fig:sdn_ex}), the forwarding devices $r_1,r_3$, and $r_2$ must be re-configured, independently.

SDN overcomes these limitations by separating the control and data-plane layers \cite{sdn}. With a centralized SDN controller, SDN applications can automatically re-configure the SDN data-plane in a timely manner. Furthermore, the devices in the data-plane may have different configuration protocols and interfaces (called the southbound interface), while the SDN controller has a single communication protocol (northbound interface) with the applications; thus, simplifying communication with heterogeneous and vendor-agnostic data-planes. One of the most popular southbound protocols is the OpenFlow protocol \cite{openflow}, where the controller gets notifications of new traffic entering the data-plane (a forwarding device as a switch or a router) when no matching rules are found. Finally, SDN-enabled forwarding devices steer the incoming network packets based on so-called flow rules installed by the SDN applications (through the controller). It is important to note that our approach is generic to any centralized routing management paradigm. Nonetheless, as SDN is the only widespread technology that provides the desired capabilities, we focus on SDN as an enabler technology.

\subsection{Definitions and notations}

First, let us consider a transportation network, where each of the links of this network has an associated link security scheme, as an example:
\begin{enumerate}[label={\arabic* -- }]
    \setcounter{enumi}{-1}
    \item No security
    \item Link layer encryption, for example, with the Encryption Control Protocol~(ECP) \cite{rfc1968}
    \item Post Quantum Cryptography~(PQC) \cite{postquantumcrypto,psc_johana}
    \item Quantum Key Distribution~(QKD) \cite{seccomqkd}
\end{enumerate}

First, we formally define the notion of security schemes and link security schemes.

\begin{definition}\label{def:secscheme} (Security Scheme) The link security schemes is a set $X$; A security scheme is an element in $X$, where a total order $\preceq$ is defined for $X$, that is for all $\alpha,\beta, \gamma\in X$:
\begin{itemize}
    \item $\alpha\preceq \alpha$ ($\preceq$ is reflexive) 
    \item $\alpha\preceq \beta$ and $\beta\preceq\gamma\implies\alpha\preceq\gamma$ ($\preceq$ is transitive)
    \item $\alpha\preceq\beta$ and $\beta\preceq\alpha\implies \alpha=\beta$ ($\preceq$ is antisymmetric)
    \item $\alpha\preceq\beta$ or $\beta\preceq\alpha$ ($\preceq$ is total)
\end{itemize}
\end{definition}

The fact that the security schemes have a defined total order is important, as there is an intrinsic sense of comparison between them. Furthermore, any two schemes can be compared and one of them is always \emph{better} than the other. Thus, for  easiness, we use the integer values associated to the schemes as in the previously presented enumeration (and the associated order relation less or equal than, denoted $\leq$). Once having the notion of link security schemes, we can proceed to define a secure network.

\begin{definition}\label{def:secnet} (Network) A secure network $N$ is a directed and weighted graph $(V,E,s)$, where:
	\begin{itemize}
			\item $V$ is a set of nodes;
			\item $E\subseteq V\times V$ is a set of edges (ordered pairs of nodes);
			\item $s: E\to X\subseteq\mathds{Z}_{+}$ is a security level function, that maps an edge to a security scheme ($x\in X$).
	\end{itemize}
\end{definition}

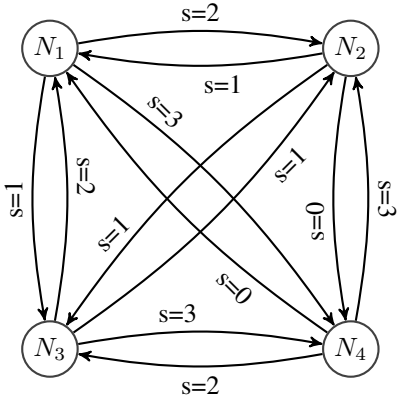
\begin{figure}
    \centering
    \begin{tikzpicture}[node distance=4cm,>=stealth',bend angle=45,auto]
    \tikzstyle{switch}=[circle,thick,draw=black!75, inner sep=0.075cm]
    \tikzstyle{host}=[circle,thick,draw=black!75,fill=black!80,text=white, inner sep=0.075cm]
    \tikzstyle{undirected}=[thick]
    \tikzstyle{directed}=[thick,->]

    \node[switch] (s1) {$N_1$};
    \node[switch, right of=s1] (s2) {$N_2$};
    \node[switch, below of=s1] (s3) {$N_3$};
    \node[switch, below of=s2] (s4) {$N_4$};
   
    \path   
            (s1)    edge[directed, bend left=10]    node[above] { s=2 }   (s2)
                    edge[directed, bend right=10]    node[above, rotate=90] {s=1}   (s3)
                    edge[directed, bend left=10]    node[above, pos=0.25, rotate=-45] {s=3}   (s4)
            (s2)    edge[directed, bend left=10]    node[below, pos=0.40] {s=1}  (s1)
                    edge[directed, bend right=10]    node[above, pos=0.75, rotate=45] {s=1}   (s3)
                    edge[directed, bend right=10]    node[above, pos=0.60, rotate=90] {s=0}   (s4)
            (s3)    edge[directed, bend right=10]    node[above, pos=0.6, rotate=-90] {s=2}   (s1)
                    edge[directed, bend right=10]    node[below, pos=0.75, rotate=45] {s=1}   (s2)
                    edge[directed, bend left=10]    node[above, pos=0.4] {s=3}   (s4)
            (s4)    edge[directed, bend left=10]    node[below, pos=0.25, rotate=-45] {s=0}   (s1)
                    edge[directed, bend right=10]    node[above, rotate=-90] {s=3}   (s2)
                    edge[directed, bend left=10]    node[below] {s=2}   (s3);
\end{tikzpicture}
    \caption{Example secure network topology}
    \label{fig:ex_topo}
\end{figure}

This network intuitively represents a resource where links have an associated level of security (0=no security, 1=ECP, 2=PostQuantum, and 3=QKD). As an example, for the network presented in Fig.~\ref{fig:ex_topo}, the model representing it, is $(V,E,s)$, where:
\begin{itemize}
    \item $V=\{N_1,N_2,N_3,N_4\}$;
    \item $E=\{(N_1,N_2),(N_1,N_3),(N_1,N_4),(N_2,N_3),(N_2,N_4),(N_3,N_4),(N_2,N_1),(N_3,N_1),(N_4,N_1),(N_3,N_2),(N_4,N_2),(N_4,N_3)\}$;
    \item $s(e)=\begin{cases}
        		0, & \text{if } e \in \{(N_4,N_1),(N_2,N_4)\}\\
        		2, & \text{if } e \in \{(N_1,N_2),(N_4,N_3),(N_1,N_3)\}\\
        		3, & \text{if } e \in \{(N_1,N_4),(N_4,N_2),(N_3,N_4)\}\\       		
				1, & \text{otherwise}
				%1, & \text{if } e \in \{(N_2,N_1),(N_2,N_3),(N_3,N_2),(N_1,N_3)\}\\
        	\end{cases}\
    $;
\end{itemize} 

Following the traditional notions in graphs, we formally consider a path in a graph as a sequence of edges, i.e., a path $p\in E^*$. As an example, the path $N_1\mapsto N_2\mapsto N_3\mapsto N_4$ is formally the sequence of edges $(N_1,N_2)(N_2,N_3)(N_3,N_4)$. As a path is a sequence, an empty path is denoted as $\epsilon$. Likewise, the concatenation of two paths $p$ and $q$ is denoted as $p\cdot q$. A \emph{simple} path, is a path that does not traverse twice any given node; for example, the path $(N_1,N_2)(N_2,N_3)(N_3,N_1)(N_1,N_4)$ is not a simple path as it traverses twice the node $N_1$. We note that, for convenience, we denote the source of an edge $e$ as $\mathbf{src}(e)$, and its destination as $\mathbf{dst}(e)$. Likewise, we denote the $i$-th edge of a path $p$ as $p_i$. We denote the first edge of a path with the index 1. Similarly, we denote the length of a path $p$ as $|p|$. It is important to note that a \emph{valid} path respects the constraint that $\forall i\in\{1,\ldots,|p|-1\} \; \mathbf{dst}(p_i)=\mathbf{src}(p_{i+1})$, i.e., the path is \emph{connected}. Finally, we denote a path $p$ starting at a node $a$ and finishing at a node $b$ (i.e., $\mathbf{src}(p_1)=a$ and $\mathbf{dst}(p_{|p|}=b$) as $p=a\to b$.

Having the definition of a network (with security levels), we focus our attention to the flows traversing such network.

\begin{definition}\label{def:sflow} (Flow) A network flow $f$ in a network $N=(V,E,s)$ is a tuple, $(o,d,h)$. $o,d\in V$, are source / origin ($o$) and destination ($d$) nodes for the flow, and $h\in\{0,1\}^k$ is the packet header (of $k$ bits) of the packets belonging to the flow, containing the common characteristics that distinguish it (from other flows). 
\end{definition}

For convenience, hereafter for a flow $f=(o',d',h')$ we use the following notations: $o(f)=o'$, $d(f)=d'$, and $h(f)=h'$. In our problem of interest, flows have associated minimal security requirements. For that reason, we define the following objects.

\begin{definition}\label{def:secreq} (Security requirement) A minimum security requirement for the header of the packets belonging to a flow $h(f)$ is a function $\mathcal{S}: \{0,1\}^k\to X\subseteq\mathds{Z}_{+}$.
\end{definition}

Having our flow and security definitions, we proceed to formally define the problem of interest.

\section{Routing under minimal security constraints}\label{sec:farsec}
\paragraph*{The flow admission and routing problem under minimal security constraints~(FARSec).} Given a network $N=(V,E,s)$, a finite set of flows $F$, and a security requirement $\mathcal{S}$, find a path for each flow (mapping solution) $\mathbb{S}: F\to E^*$, such that: i) the paths are simple and valid for the flows and network; ii) the required security level of the allocated flows is lesser or equal than the security value of each of the links in the assigned paths or the empty path if the latter is not possible. These two conditions are hereafter referred to as \emph{the properties of a FARSec solution}.

As an example, consider the set of flows as shown in Table~\ref{tab:ex_flows} and the network shown in Figure~\ref{fig:ex_topo}. The solution $\mathbb{S}=\{(0001,(N_1,N_4)(N_4,N_2)),(0010,\epsilon),(0011,(N_3,N_4)(N_4,N_2)), (0100,(N_4,N_3)(N_3,N_1))\}$ is a FARSec solution. Note that, the flow with header $0010$ is assigned an empty path as all available paths from the source ($N_2$) to the destination ($N_4$) traverse links which have lower security level than the required one ($2$). In this case, in order not to compromise the security of the running application (with header $0010$) the traffic is dropped. 

\begin{table}[!htb]
	\centering
	\begin{tabular}{|c|c|c|c|c|c|}
		\hline
		\textbf{header} & \textbf{source} 	& \textbf{destination} & \textbf{min. sec.} \\ \hline
		0001           & $N_1$         & $N_2$ & 3  \\ \hline
		0010           & $N_2$         & $N_4$ & 2  \\ \hline
		0011           & $N_3$         & $N_2$ & 1  \\ \hline
		0100           & $N_4$         & $N_1$ & 2  \\ \hline
	\end{tabular}
	\caption{Example flows}
	\label{tab:ex_flows}
\end{table}

\subsection{Computing the solution for the FARSec problem}
In order to compute a solution for a given FARSec problem instance, different strategies are possible. The naive option is to go through each of the flows to allocate and recursively search for a path starting at the source node whose edges have greater or equal security level as the requested one. However, in the worst case, exponential number of paths should be checked. Indeed, a better strategy is to search for one path for each source / destination pair; that is the path with the maximal minimum security level. Subsequently, for each requested flow, if the path with the maximal minimum security level has no edge with a level less than the required security level, this path can be assigned to the flow; otherwise, the flow must be dropped (the empty path must be assigned). The latter strategy reduces the problem to computing all-pairs \emph{widest paths}. There are many possibilities how to compute all-pairs widest paths. Some of them are better for dense graphs, and others for sparse ones. Without any knowledge of which sort of input graphs are to be treated, a good option is to use a modified version of the simple and efficient ($\Theta(|V|^3)$) algorithm known as the Floyd-Warshall algorithm, originally proposed by Bearnand Roy\cite{allpairs} (later, independently discovered by Robert Floyd and Stephen Warshall, see for example \cite{floydallpairs}). 

The difference between computing the shortest paths and the widest paths can be easily explained via the following recurrences. First, the shortest path from a source node $a$ to a destination node $b$ can be computed with the recurrence $SP(a,b)= \mathbf{min}\{SP(a,u)+s((u,b))|(u,b)\in E\}$, where $s$ is the weight function of the graph (in our case the security level). This formulation is known and can be used to implement many algorithms such as Dijkstra's algorithm or Roy's algorithm for all pairs. The recurrence to describe the widest path ($a\to b$) is as follows: $WP(a,b)=\mathbf{max}\{\textbf{min}\{WP(a,u),s((u,b))\}|(u,b) \in E\}$. Although these recurrences may look very different, in fact, there are only two different operators. Namely, $\mathbf{min}$ shortest paths is replaced by $\mathbf{max}$ in widest paths and $+$ in shortest paths  is replaced by $\mathbf{min}$. Indeed, the widest paths search to maximize the minimal security level. We present the modified Roy's algorithm to compute the all-pairs widest paths in Algorithm~\ref{algo:widest} while Algorithm~\ref{algo:FARSec} presents how to solve the FARSec problem by mapping each flow to a path, whenever possible.
\begin{algorithm}[!htb]
    %\small
    \SetKwInOut{Input}{input}\SetKwInOut{Output}{output}\SetKw{KwBy}{by}
    \DontPrintSemicolon
    \Input{A secure network graph $(V,E,s)$}
    \Output{$P$, the most secure paths matrix where $P_{i,j}$ contains the secure most path from the node $i\to j$ for all $i,j\in V$}
    \textbf{Step 0}: Initialize $\omega$, $P$ and $\Pi$ as a $|V|\times|V|$ integer matrices and set $P_{i,j}\leftarrow\epsilon$, $\Pi_{i,j}\leftarrow$\textsc{NULL}, and $\omega_{i,j}\leftarrow0$ for all $i,j\in\{1,\ldots,|V|\}$\tcp{$\omega$ represents the smallest security level of the path, $\Pi$ the next hop in the path, and $P$ the final paths}
    \textbf{Step 1}: \ForEach{$(u,v)\in E$}
    {
        Set $\omega_{u,v}\leftarrow s((u,v))$\tcp{The existing security level of the link}
        Set $\Pi_{u,v}\leftarrow v$\;
    }
    \textbf{Step 2}: \ForEach{$v\in V$}
    {
        Set $\omega_{v,v}\leftarrow \infty$\;
        Set $\Pi_{v,v}\leftarrow v$
    }
    \tcp{Start the standard Floyd-Warshall algorithm}
    \textbf{Step 3}: \For{$k\leftarrow 1; k \leq |V|; k\leftarrow k+1$}
    {
        \For{$i\leftarrow 1; i \leq |V|; i\leftarrow i+1$}
        {
            \For{$j\leftarrow 1; j \leq |V|; j\leftarrow j+1$}
            {
                %original if dist[i][j] > dist[i][k] + dist[k][j] then
                \If{$\omega_{i,j} < \mathbf{min}\{\omega_{i,k},\omega_{k,j}\}$}
                {
                    Set $\omega_{i,j}\leftarrow \mathbf{min}\{\omega_{i,k},\omega_{k,j}\}$\;
                    Set $\Pi_{i,j}\leftarrow\Pi_{i,k}$\;
                }
            }
        }
    }
    \tcp{At this point we must reconstruct the paths}
    \textbf{Step 4}: \For{$i\leftarrow 1; i \leq |V|; i\leftarrow j+1$}
    {
        \For{$j\leftarrow 1; j \leq |V|; j\leftarrow j+1$}
        {
            \If{$\Pi_{i,j} \not=$  \textsc{NULL}}
            {
                $curr \leftarrow i$\;
                \While{$curr\not= j$}
                {
                    Set $P\leftarrow P\cdot (curr, \Pi_{curr,j})$\;
                    Set $curr\leftarrow \Pi_{curr,j}$\;
                }
                  
            }
            
        }
    }
    \textbf{Step 5}: \Return{$P$}
    \caption{All-pairs widest paths}\label{algo:widest}
\end{algorithm}

\begin{algorithm}[!htb]
    %\small
    \SetKwInOut{Input}{input}\SetKwInOut{Output}{output}\SetKw{KwBy}{by}
    \DontPrintSemicolon
    \Input{A secure network graph $(V,E,s)$, a set of flows $F$, and a security requirement 
    $\mathcal{S}$}
    \Output{A mapping $\mathbb{S}$, from $F$ to $E^*$}
    \textbf{Step 0}: Set $\mathbb{S}\leftarrow\emptyset$, and initialize $L$ as a $|V|\times|V|$ integer matrix, where $L_{i,j}=\infty$, $\forall i,j\in\{1,\ldots,|V|\}$ \tcp{$L$ contains the bottleneck security level for each path.}
    \textbf{Step 1}: Use Algorithm~\ref{algo:widest} to compute $P$\;
    \textbf{Step 2}:  \For{$i\leftarrow 1; i \leq |V|; i\leftarrow k+1$}
    {
        \For{$j\leftarrow 1; j \leq |V|; j\leftarrow j+1$}
        {
            \For{$k\leftarrow 1; k \leq |P_{i,j}|; k\leftarrow k+1$}
            {
                \If{$s({P_{i,j}}_k) < L_{i,j}$}
                {
                    Set $L_{i,j}\leftarrow s({P_{i,j}}_k)$\tcp{Update the real bottleneck value}
                }
            }
        }
    }
    \tcp{Do the main mapping here}
    \textbf{Step 3}: \ForEach{$f\in F$}
    {
        \If{$\mathcal{S}(h(f)) \leq L_{o(f),d(f)}$}
        {
            Set $\mathbb{S}\leftarrow\mathbb{S}\cup\{(f,P_{o(f),d(f)})\}$\;
        }
        \Else
        {
            Set $\mathbb{S}\leftarrow\mathbb{S}\cup\{(f,\epsilon)\}$\tcp{If there is no secure enough path, drop the traffic}
        }
    }
    \textbf{Step 4}: \Return{$\mathbb{S}$}
    \caption{FARSec solution}\label{algo:FARSec}
\end{algorithm}

To assess the correctness of Algorithm~\ref{algo:FARSec} we first consider that the algorithm always terminates as there are loops only of finite length. Additionally, as the widest paths computation ensures that there does not exist a path which has a lager minimal security path and this path is assigned to the flow only if the requirement is lesser or equal than the bottleneck of this path (otherwise the empty path is assigned), the following statement holds.

\begin{proposition}
Given a secure network graph $(V,E,s)$, a set of flows $F$, and a security requirement $\mathcal{S}$, $\mathbb{S}$ computed by Algorithm~\ref{algo:FARSec} holds the properties of a FARSec solution, i.e.,
\begin{enumerate}
    \item $\forall f=(o,d,h)\in F\; ((\mathbb{S}(o,d)\not=\epsilon) \implies ( (\forall i\in\{1,\ldots,|\mathbb{S}(o,d)|-1\} \; \mathbf{dst}(\mathbb{S}(o,d)_i)=\mathbf{src}(\mathbb{S}(o,d)_{i+1})) \wedge (\forall x\in\{1,\ldots,|\mathbb{S}(o,d)|-1\}\;(\nexists y\in\{x+1,\ldots,|\mathbb{S}(o,d)|\}\;(\mathbf{src}(\mathbb{S}(o,d)_x)=\mathbf{dst}(\mathbb{S}(o,d)_y)))) ) )$ (paths are simple valid if not the empty path)
    \item $\forall f=(o,d,h)\in F\; ((\mathbb{S}(o,d)\not=\epsilon) \implies (\forall i\in\{1,\ldots,|\mathbb{S}(o,d)|\} \mathbb{S}(o,d)_i \geq \mathcal{S}(h)))$ (assigned paths respect the minimal security constraints if assigned)
    \item  $\forall f=(o,d,h)\in F\; ((\mathbb{S}(o,d)=\epsilon) \implies (\nexists p\in E^* ((p\not=\epsilon) \wedge (\mathbf{src}(p_1)=o) \wedge ((\mathbf{dst}(p_{|p|})=d) \wedge (\forall i\in\{1,\ldots,|p|-1\} \; \mathbf{dst}(p_i)=\mathbf{src}(p_{i+1})) \wedge (\forall x\in\{1,\ldots,|p|-1\}\;(\nexists y\in\{x+1,\ldots,|p|\}\;(\mathbf{src}(p_x)=\mathbf{dst}(p_y))) ) \wedge (\forall j\in\{1,\ldots,|p|\} p_j \geq \mathcal{S}(h)) ) ) )$ (there does not exist a valid and simple path for the flow that respects the security constraints if the path is not assigned by the algorithm)
\end{enumerate}
\end{proposition}

It is good to have security guarantees in order to ensure dependability on such critical systems. However, an interesting aspect to consider is the computational complexity of the overall proposed approach. Differently, from the naive possibly exponential computation, our proposed approach has a polynomial worst-case guarantee. 

\subsubsection{Complexity analysis of the proposed FARSec solution}
It is easy to see that the FARSec solution (as shown in Algorithm~\ref{algo:FARSec}) has a $\mathcal{O}(|V|^4+|F|)$ time complexity, as for each pair of nodes the assigned path (is simple and thus) has at most $\mathcal{O}(|V|^2)$ edges (hence the $|V|^2$ term), and then all flows are visited once. We note that the bottleneck weights can be obtained directly from the computation of the widest paths. However, when implementing the approach it is often desirable to have dedicated functions for code re-usability and testability purposes. One function computing only the widest paths and the other computing a bottleneck weights for a path. In case of computing both combined, the running time would be $\Theta(|V|^3+|F|)$, which is not such a significant improvement for a small number of nodes. Furthermore, for networks where nodes have relatively short paths (for example double-stared networks, which are common network architectures), the time complexity of computing the bottleneck of the path gets dramatically decreased (as there does not exist a simple and valid path with $\mathcal{O}(|V|^2)$ nodes in such topologies). For our particular instances of interest both of the aforementioned conditions are true, i.e., the number of nodes is rather small and the paths are short. 

\section{Complexity and scalability evaluation}\label{sec:experiments}
\subsection{Experimental evaluation}
\begin{figure}[!ht]
    \centering 
    \includegraphics[width=\columnwidth]{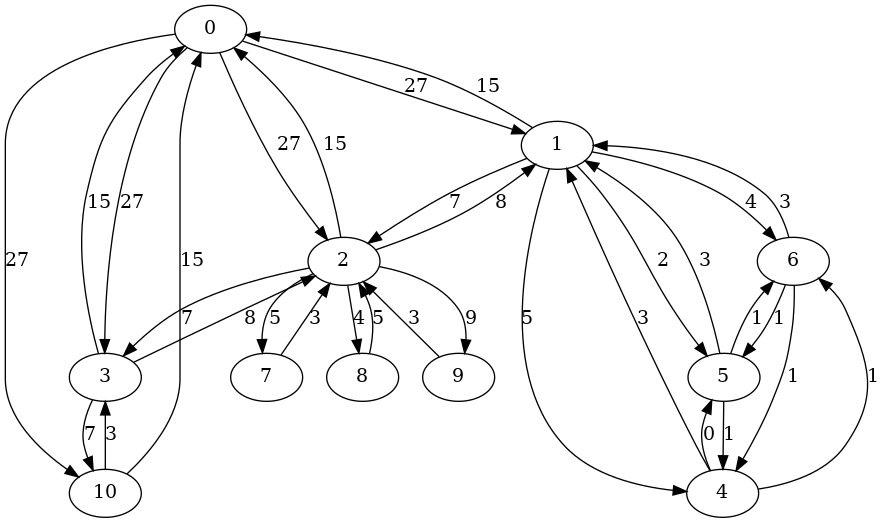}
    \caption{Example tested topology}
    \label{fig:ex_test_topo}
\end{figure}

In order to showcase the effectiveness of the proposed approach, semi-random instances were generated. Indeed, in order to ``test'' different topologies at once, a so-called double star (a rooted tree with three levels) topology has been used as a base. On the second level, nodes are interconnected to their immediate neighbors, forming a so-called bus topology. Finally, in the third level, one of the nodes has a fully connected component, forming a so-called mesh network. As an example, a randomly generated topology for 11 nodes corresponds to the one shown in Figure~\ref{fig:ex_test_topo}. The choice of the security levels is also done in a semi-random manner. The security level from the root to the first level were chosen as a random integer between 0 and 30; respectively, the security levels from the second level were chosen as a random integer between 0 and 10;  finally, the links at the third level were chosen as a random between 0 and two. The number of flows is set to be high, 64 times the sum of all outgoing edge weights. With these settings, we generate the secure network and the flow instances as inputs for the proposed solution.

\paragraph{Experimental setup.} The experiments were performed in a commodity server with 48 AMD cores, and 128MB of RAM, running a Linux 20.04 LTS. Random instances were generated ranging from two to 50 nodes. The running time for solving the FARSec problem w.r.t. to the size of the network ($|V|$) is shown in Figure~\ref{fig:farsec_rt}. Similarly, the running time for solving the FARSec problem w.r.t. to the size of the flows ($|F|$) is shown in Figure~\ref{fig:farsec_rt_f}.  As can be seen, the running time of the FARSec algorithm is quite low (and generally depends on the dominating term, which seems to be the number of flows). For generating and visualizing the random instances, a web service has been created, at the time of this writing, it is accessible through the following URL: \href{http://www.ailab.airbus.com/TRUSTCOM23/Frontend/?size=S}{\url{http://www.ailab.airbus.com/TRUSTCOM23/Frontend/?size=S}}, where ``S'' must be replaced by a size, for example for an 11 node instance \href{http://www.ailab.airbus.com/TRUSTCOM23/Frontend/?size=11}{\url{http://www.ailab.airbus.com/TRUSTCOM23/Frontend/?size=11}}.

\begin{figure}[!htb]
    \centering
    \begin{tikzpicture}[scale=1]
    \begin{axis}[
        xlabel=Instance size ($|V|$),
        ylabel=Time (s),
        legend style={at={(0.02,0.98)}, anchor=north west},
        ytick = {0.01, 0.02, 0.03, 0.04, 0.05}
        ]
    
    \addplot[smooth,mark=*, mark size=0.8pt, black] plot coordinates {
		(2, 0.0016)
		(3, 0.0035)
		(4, 0.0046)
		(5, 0.0053)
		(6, 0.0073)
		(7, 0.0098)
		(8, 0.011)
		(9, 0.0093)
		(10, 0.0074)
		(11, 0.0093)
		(12, 0.011)
		(13, 0.012)
		(14, 0.014)
		(15, 0.015)
		(16, 0.012)
		(17, 0.012)
		(18, 0.014)
		(19, 0.016)
		(20, 0.017)
		(21, 0.019)
		(22, 0.024)
		(23, 0.026)
		(24, 0.027)
		(25, 0.021)
		(26, 0.02)
		(27, 0.022)
		(28, 0.024)
		(29, 0.028)
		(30, 0.03)
		(31, 0.032)
		(32, 0.033)
		(33, 0.034)
		(34, 0.038)
		(35, 0.04)
		(36, 0.027)
		(37, 0.027)
		(38, 0.03)
		(39, 0.032)
		(40, 0.034)
		(41, 0.036)
		(42, 0.037)
		(43, 0.048)
		(44, 0.05)
		(45, 0.052)
		(46, 0.037)
		(47, 0.039)
		(48, 0.039)
		(49, 0.034)
		(50, 0.034)
    };
    
    \end{axis}

\end{tikzpicture}
    \caption{FARSec running time experiments (w.r.t. $|V|$)}
    \label{fig:farsec_rt}
\end{figure}

\begin{figure}[!htb]
    \centering
    \begin{tikzpicture}[scale=1]
    \begin{axis}[
        xlabel=Instance size ($|F|$),
        ylabel=Time (s),
        legend style={at={(0.02,0.98)}, anchor=north west},
        ytick = {0.01, 0.02, 0.03, 0.04, 0.05}
        ]

    \addplot[smooth,mark=*, mark size=0.8pt, black] plot coordinates {
        (1922,0.0016)
        (4163,0.0035)
        (5380,0.0046)
        (6213,0.0053)
        (8454,0.0073)
        (11271,0.0098)
        (13192,0.011)
        (10825,0.0093)
        (8842,0.0074)
        (10955,0.0093)
        (12812,0.011)
        (14541,0.012)
        (15886,0.014)
        (17231,0.015)
        (13456,0.012)
        (13841,0.012)
        (15954,0.014)
        (17811,0.016)
        (19540,0.017)
        (21397,0.019)
        (27350,0.024)
        (29079,0.026)
        (30808,0.027)
        (22809,0.021)
        (23130,0.02)
        (25371,0.022)
        (27740,0.024)
        (31389,0.028)
        (33374,0.03)
        (35487,0.032)
        (37216,0.033)
        (38945,0.034)
        (43618,0.038)
        (45923,0.04)
        (30948,0.027)
        (31461,0.027)
        (34086,0.03)
        (35687,0.032)
        (38248,0.034)
        (40297,0.036)
        (42346,0.037)
        (54251,0.048)
        (56748,0.05)
        (59245,0.052)
        (42158,0.037)
        (43311,0.039)
        (44464,0.039)
        (37617,0.034)
        (36210,0.034)
    };
    
    \end{axis}

\end{tikzpicture}
    \caption{FARSec running time experiments (w.r.t. $|F|$)}
    \label{fig:farsec_rt_f}
\end{figure}

\section{An open source implementation of the proposed approach}\label{sec:imp}
\begin{figure}[!htb]
    \centering 
    \includegraphics[width=\columnwidth]{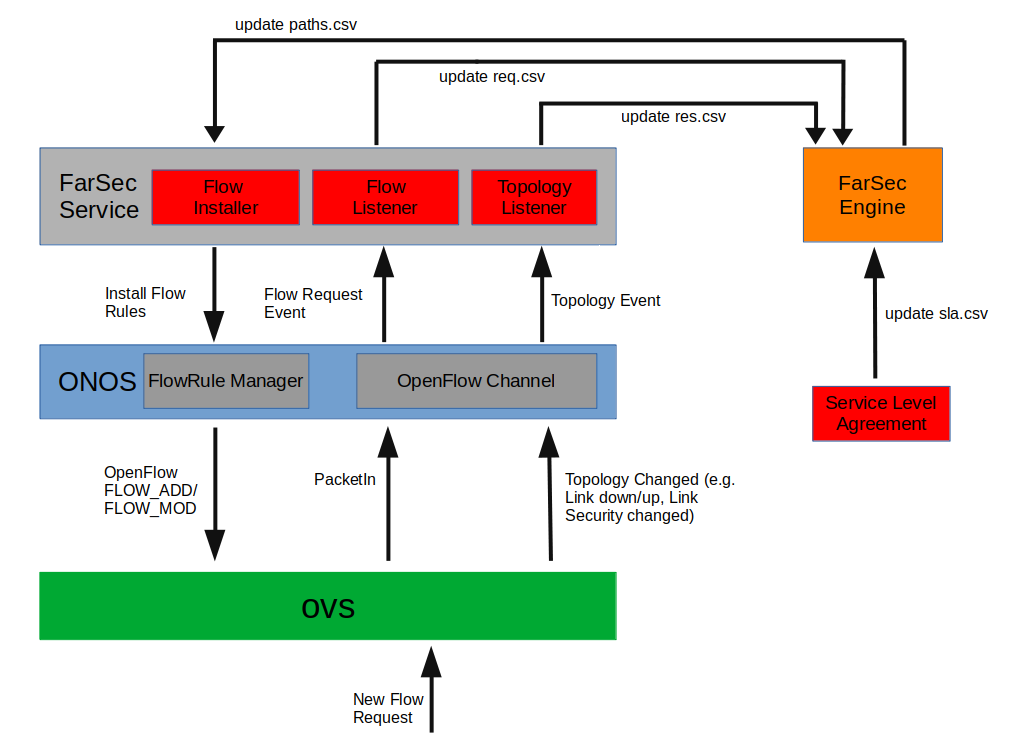}
    \caption{FARSec architecture}
    \label{fig:farsec_arch}
\end{figure}

We implemented the proposed solution as an SDN service using the ONOS \cite{onos} controller. ONOS is an Open Networking Foundation~(ONF) project and one of the two most widely used open-source SDN controllers. The complete solution is comprised of two software components i) The \verb|FARSec ONOS| service, written in Java which uses the ONOS Northbound API, and ii) the \verb|FARSec engine| service written in C++ which implements the algorithm described in Section~\ref{sec:farsec}. In general, the controller receives either new flow events or link change events (e.g., change in security level), and forwards the information to the \verb|FARSec ONOS| service. The service decides on the pertinent action (e.g., change the topology information) and forwards it to the \verb|FARSec engine| service. The latter responds with the expected mapping configuration, and the \verb|FARSec ONOS| service then pushes the rules via the controller. A graphic view of the overall process is shown in Figure~\ref{fig:farsec_arch}. A detailed description of each component is further presented.

\subsection{FARSEC Engine}\label{sec:farsec-engine}
The \verb|FARSec engine| service takes three inputs which in the current version of our software are given as CSV files. The resources input (an example in Figure~\ref{fig:res_csv_ex}) contains a list of all the active links in the topology and their corresponding security level. The requests input (an example in Figure~\ref{fig:req_csv_ex}) contains the list of all the flows that need to be distributed in the network. Finally, the Service Level Agreement~(SLA) input (an example in Figure~\ref{fig:sla_csv_ex}) contains the security requirements that need to be respected when calculating the paths. The engine computes the solution of the FARSec instance, and outputs the mapping between the flows and the associated paths.

\begin{figure}[!htb]
\begin{lstlisting}[language=bash,basicstyle={\scriptsize}, backgroundcolor=\color{gray!20!white}]
Source,Destination,Security
s1,s2,1
s1,s3,3
s2,s3,2
\end{lstlisting}
\caption{Example resources CSV file}
\label{fig:res_csv_ex}
\end{figure}

\begin{figure*}[!htb]
\begin{lstlisting}[language=bash,basicstyle={\scriptsize}, backgroundcolor=\color{gray!20!white}]
FlowID,Source,Destination,Header
600309120,s1,s3,450000140000000000013597c0a80101c0a803010008fff700000000000000000000000000000000
142719360,s3,s1,450000140000000000013597c0a80301c0a801010008fff7000000000000000000000000000000000
\end{lstlisting}
\caption{Example Requests CSV file}
\label{fig:req_csv_ex}
\end{figure*}

\begin{figure*}[!htb]
\begin{lstlisting}[language=bash,basicstyle={\scriptsize}, backgroundcolor=\color{gray!20!white}]
Protocol,SourceAddress,DestinationAddress,DSCP,SourcePortMin,SourcePortMax,DestinationPortMin,DestinationPortMax,MinSec
UDP,0.0.0.0/0,0.0.0.0/0,0,0,65535,5000,5005,2
TCP,192.168.1.0/24,192.168.2.0/24,0,65535,22,22,4
\end{lstlisting}
\caption{Example SLA CSV file}
\label{fig:sla_csv_ex}
\end{figure*}

\subsection{FARSEC ONOS Service}\label{sec:farsec-service}
The resources and requests CSV files that are consumed by the \verb|FARSec engine| service for the path computation are generated reactively by the \verb|FARSEC ONOS service|. FARSEC is comprised of three components. The resources file which contains the topology is updated whenever the service receives an event from the network. Examples of events are the appearance / disappearance of a device (Open vSwitch\cite{ovs}), link up / link down or a link whose security level has changed. The Topology Listener (object from ONOS API) component that is in charge of updating the resources file and feeding it to the engine. 

A separate component called the Flow Listener reacts on flow events. Every time an Open vSwitch device receives a packet for which it does not have a rule installed on its tables it sends an OpenFlow \verb|PacketIn| \cite{openflow} message. Upon reception of a \verb|PacketIn| message by the controller, the Flow Listener component of the \verb|FARSec ONOS| service gets the packet header and infers the source and destination switches from the respective source and destination IP addresses. Packets are generated by network hosts. Every host has an attachment point to a switch in the topology which is known to the controller so the latter can infer the source / destination switches. The service then updates the request CSV file with the new flow and passes it to the \verb|FARSec engine| service for path calculation. As a reminder, Figure~\ref{fig:farsec_arch} illustrates the whole architecture. 

\subsection{SLA Component}\label{sec:sla-component}
The SLA component is a CSV file that for the current FARSec version needs to be updated manually. The SLA CSV file is a list of security requirements. Each entry contains some IP header fields (see detailed description in Table~\ref{tab:sla_fields}) that can be matched against every incoming flow, and a corresponding minimal security level requirement for the matched flows. If a flow does not match any of the SLA entries it is given the default minimum security requirement of 0. 
\begin{table*}[!htb]
	\centering
	\begin{tabular}{|c|c|c|c|}
		\hline
		\textbf{Field} & \textbf{Description} 	& \textbf{Example Values} \\ \hline
		Communication Type & The IP protocol & UDP,TCP,ICMP  \\ \hline
		Source/Destination Address & The source/destination IP address & 0.0.0.0/0, 192.168.1.4/24  \\ \hline
		DSCP & The TOS field in the IP header & 0,1,8,40,46  \\ \hline
        Source/Destination Port & The source/destination port of the flow (can also be a range) &  80, 22, 5000-5010 \\ \hline
        MinSec & The minimum security requirement & 1,2,3,4 \\ \hline
	\end{tabular}
	\caption{SLA fields}
	\label{tab:sla_fields}
\end{table*}

Consider the SLA CSV file shown in Figure~\ref{fig:sla_csv_ex}. The semantic of this file is as follows. All UDP flows from any source, destination IP and any source port that have a destination port between 5000 and 5005 have a minimum security requirement of 2. Respectively, all TCP flows from the subnet 192.168.1.0/24 to subnet 192.168.2.0/24 that have 22 as the destination port have a security requirement of 4. In both examples, DSCP is not used so it is set to zero.

\subsection{Video Demonstrator}\label{sec:demo}
\begin{figure}[!ht]
    \centering 
    \includegraphics[width=\columnwidth]{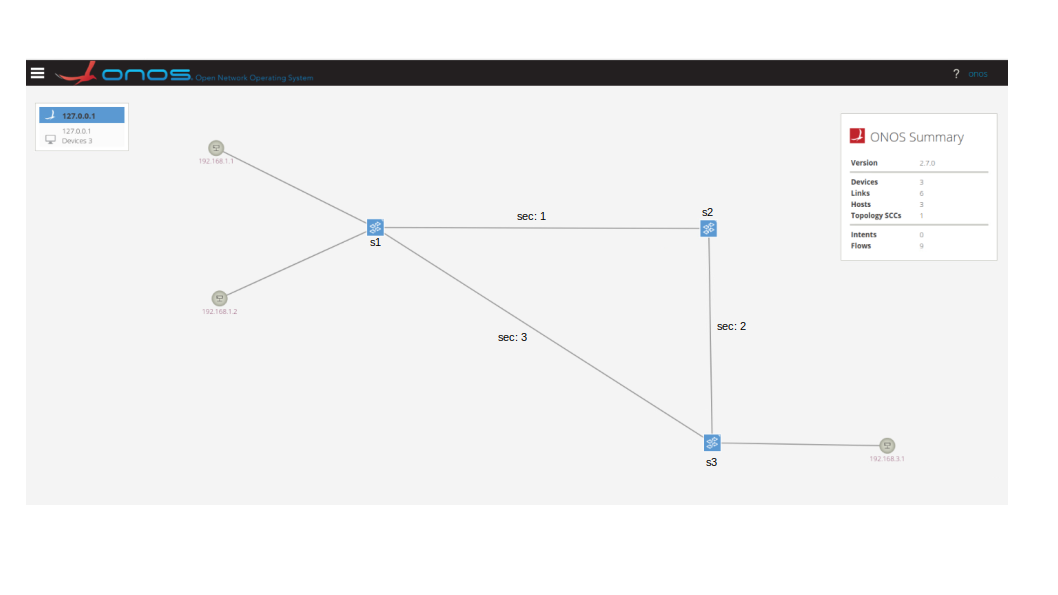}
    \caption{Network topology from ONOS controller}
    \label{fig:farsec_onos_topo}
\end{figure}

At the time of this writing, the source code is not publicly available but, it can be made available for research collaborations (the interested reader can contact the corresponding author). However, in order to better showcase the overall functionality a video can be found here: \href{http://www.ailab.airbus.com/TRUSTCOM23/Frontend/farsec-demo.mp4}{\url{http://www.ailab.airbus.com/TRUSTCOM23/Frontend/farsec-demo.mp4}}.
To better explain the functionality and to easily assess the expected results, we use the simple topology presented in Figure~\ref{fig:farsec_onos_topo}. Two UDP video flows are initiated from h1 (192.168.1.1, attached to switch S1) and are destined to h3 (192.168.3.1, attached to s3). Initially all links have a security tier of 4 and the flows have the minimum security requirement of 3. We then modify the link security tiers as well as the flows' security requirements and we observe how the FARSec application correctly performs rerouting and flow admission respecting the current security requirements of the traffic.

%\begin{figure}[!ht]
%    \centering 
%    \includegraphics[width=\columnwidth]{figures/farsec_demo.png}
%    \caption{FARSec Demonstrator}
%    \label{fig:farsec_demo}
%\end{figure}

\section{Related Work}\label{sec:relwork}
Constraint-based routing (CBR) is an active area of research. Several works in the literature have dealt with the problem of optimizing routes based on a number of link metrics such as available bandwidth, delay, jitter, packet loss, hop count or other types of costs. Based on the type and subset of metrics as well as the type of the optimization problem a number of works have been proposed. Goyal and Hjalmytsson \cite{qos-routing} tackle the bandwidth-bounded problem whereas \cite{qos-routing-inaccurate}, \cite{distributed-route} deal with the delay-bounded path search. Some of the examples in \cite{qos-multimedia} also deal with two-metric constraint optimizations like bandwidth-bounded/delay-bounded or bandwidth-optimized/delay-optimized. The Extended Bellman-Ford (EBF) and Extended Djikstra Shortest Path (EDSP) algorithms presented in \cite{qos-overview} are both tested on the bandwidth-bounded/cost-bounded problem. In the same work a number of constraint-based routing algorithms are presented along with their computational complexities. Ma and Steenkiste \cite{pathbwguarantees} propose a modified Bellman-Ford algorithm for a multi-constraint problem. The  Heuristic Multi-Constrained Optimal Path (HMCOP) algorithm proposed by Korkmaz and Krunz in \cite{multi-constrained} also attempts to find feasible paths that satisfy a set of constraints while maintaining high utilization of network resources. More recent works have also explored QoS-routing over SDN networks. \cite{qos-prioritazation-sdn} and \cite{realizing-qos-sdn} mainly focus on the optimization of bandwidth while OpenQoS \cite{openqs} and VSDN \cite{vsdn} use the hop-count and delay as QoS costraints. AmoebaNet \cite{AmoebaNet} also uses the bandwidth constraint along with a variant of Djikstra's algorithm for path computation. Finally, QoSS \cite{qoss} uses the usual QoS metrics along with a security level of nodes to compute optimal end-to-end paths. However, this work focuses more on maximizing the throughput of applications and does not perform flow admission.

To the best of our knowledge, in the literature, there does not exist a solution for the priority and admission of flows under minimal security constraints be that in a centralized (SDN) architecture or not. 

\section{Conclusion}\label{sec:conclusion}
In this paper, we have proposed a constrained-based routing algorithm where the metric of interest is the link's security level. The algorithm searches for a feasible path for each incoming flow that respects the latter's minimum level of security requirement. Experimental evaluation with different topologies comprising variable link security levels showcases the performance and pertinence of the proposed approach. The proposed solution runs in polynomial w.r.t. to the size of the network and number of flows. Finally, we have implemented the FARSec approach as a northbound SDN service using the ONOS controller. The service is in charge of monitoring the network, exporting necessary information as input to the FARSec algorithm and installing the paths calculated by the latter.

As for future work, we plan to explore the routing and admission problem with several constraints (e.g., security and capacity, security and latency etc.). Moreover, it is interesting to consider how different parameters affect the hardness of the problem.

\bibliographystyle{IEEEtran}
\bibliography{IEEEabrv,references}
\end{document}